\documentclass[a4paper]{jpconf}
\usepackage{graphicx}
\begin{document}
\title{Exact Chiral Invariance at Finite Density on Lattice }

 \author{Rajiv V. Gavai}
 
 \address{Department of Theoretical Physics, Tata Institute of Fundamental
 Research, \\ Homi Bhabha Road, Mumbai 400005, India}
 
 \ead{gavai@tifr.res.in}
 
 \author{Sayantan Sharma}
 
 \address{Fakult\"at f\"ur Physik, Universit\"at Bielefeld, D-33615
 Bielefeld, Germany}
 
 \ead{sayantan@physik.uni-bielefeld.de}

\begin{abstract}
A new lattice action is proposed for the overlap Dirac matrix with
nonzero chemical potential.  It is shown to preserve the full chiral 
invariance for all values of lattice spacing exactly. It is further 
demonstrated to arise in the domain wall formalism by coupling the 
chemical potential count only to the physically relevant wall modes. 

\end{abstract}

\section{Introduction}

A fundamental aspect of the phase diagram of Quantum Chromo Dynamics (QCD) in
the $T$-$\mu_B$ plane is the critical point, where $T$ and $\mu_B$ denote the
the temperature and baryonic chemical potential respectively.
Theoretical~\cite{fk,gg1,schmidt} as well as experimental searches for locating
it are currently going on~\cite{mohanty}.  Its discovery would be exciting in
many ways. Apart from becoming a new milestone in our understanding of the
nature of the strongly interacting matter, it would also be unique compared  to
the other known phase diagrams in the way theory and experiment compliment each
other in locating the critical point in it.   It is desirable to predict the
existence of the QCD critical point (or rule it out) starting from first
principles,  employing only the QCD Hamiltonian.  Lattice QCD is the best
suited tool for such an exploratory exercise in view of its success in other
non-perturbative aspects.  

The prime reason for expecting the critical point are the chiral symmetries QCD
has due to the quark mass spectra.  Our world  has two light quarks,  having
masses much smaller than $\Lambda_{QCD}$, and a moderately heavy quark, making
it a `2+1' flavour QCD.  All attempts to explore nonzero $\mu_B$ on the lattice
use the staggered quarks.  They have a $U(1)_V \times U(1)_A$  chiral symmetry
on the lattice, and a corresponding order parameter, the chiral condensate.
Staggered quarks, however, break the flavour and spin symmetry on the lattice,
and have no flavour singlet axial anomaly.  In the continuum limit of $a \to 0$,
they have 4 quark `flavours' of the same mass. The `2+1' flavour QCD is
approximated by taking square root and fourth root of the staggered quark
determinant.  Whether this can restore the correct flavour symmetry as well as
the chiral anomaly in the continuum limit has been a controversial issue.
Moreover, it has been argued the `rooting problem' becomes even worse
~\cite{gss} at nonzero $\mu_B$.  Since the QCD critical point is
expected~\cite{rw} to exist if only if one has two light flavours, and the
flavour singlet anomaly is mildly temperature dependent~\cite{piswil}, as in
our world, it appears desirable to improve upon them.   The overlap Dirac
fermions~\cite{Neu1}, or the closely related domain wall
fermions~\cite{kaplan}, offer such a possibility to improve.  Indeed, the
overlap quarks have all the symmetries of the continuum QCD  and also have an
index theorem~\cite{hln} as well, raising the hope that even the anomaly
effects could be well treated.   Unfortunately, adding the chemical potential
turns out to be nontrivial for them.  Bloch and Wettig~\cite{bw} made a
proposal to do so but it violates~\cite{bgs} the exact chiral invariance on
lattice as does the simple addition~\cite{gs} of a baryon number term.    In
this talk I report on our alternative action~\cite{gs12} which does have exact
and continuum-like chiral invariance on lattice for any value of the lattice
spacing and any chemical potential.  Its chiral order parameter permits, in
principle, the task of mapping out the $T$-$\mu_B$ phase diagram, assuming that
the algorithmic developments can handle the fermion sign problem well.

\section{Exact Chiral Invariance for $\mu \ne 0$}

We begin by noting that the massless continuum QCD action can be written in a 
form where the chiral symmetry is manifest in terms of the fields appearing 
in the action and ask \cite{gs12} if this can be imitated on lattice.  The 
continuum QCD action at finite density in an explicit chiral invariant form is
\begin{equation}  
\nonumber 
S_{QCD} = \int d^3x~d\tau [ \bar \psi_L ({\not}D + \mu \gamma^4) \psi_L + 
\bar \psi_R ({\not}D + \mu \gamma^4)  \psi_R - F^{\mu \nu} F_{\mu \nu} /4] ~, 
\label{eqn:qcdlr}
\end{equation} 
where $\psi_L = (1-\gamma_5)\psi/2 $ and $\psi_R = (1+\gamma_5)\psi/2 $ with $
\bar \psi_L = \bar \psi (1+\gamma_5)/2 $ and $\bar \psi_R = \bar \psi
(1-\gamma_5)/2$. 
The second term in the action is for gluons and will play no role in our
discussion below.  We assume that some usual convenient lattice form for it has
been chosen and focus on the lattice fermionic action only.  We propose that
addition of chemical potential on the lattice be done in such an explicit chiral
symmetry preserving manner as well, opting therefore for the overlap quarks on
the lattice.  

The overlap quarks have all the symmetries of the continuum QCD  but also have a
nonlocal action:   
\begin{equation} 
\label{eqn:ovelp1} 
S_F = \sum_{n,m}  \bar \psi_n~a D_{ov,nm}~\psi_m  ~, 
\end{equation} 
where the sum over $n$ and $m$ runs over all the space-time lattice sites, $a$
is the lattice spacing, and the overlap Dirac matrix $D_{ov}$ is defined by
$aD_{ov}=1+\gamma_5 sgn (\gamma_5 D_W)$.  {\em sgn} denotes the sign function. 
$D_W$ is the standard Wilson-Dirac matrix on the lattice but with a negative
mass term $M\in(0,2)$:
\begin{equation} 
\label{eqn:Dwil} 
a D_W(x,y) = (4-M)\delta_{x,y} - \sum_{i=1}^{4}
[U^{\dagger}_{i}(x-\hat{i})\delta_{x-\hat{i},y}\frac{1+\gamma_{i}}{2}
+\frac{1-\gamma_{i}}{2}U_{i}(x)\delta_{x+\hat{i},y}]~.
\end{equation} 
The overlap Dirac matrix satisfies Ginsparg-Wilson relation~\cite{wil},  
$\{\gamma_5, D\} = a D \gamma_5 D $ and has exact chiral symmetry on lattice. 
The corresponding infinitesimal chiral transformations~\cite{Lues} 
\begin{equation}
\label{eqn:chrLR} 
\delta \psi = i \alpha  \gamma_5(1 - a D_{ov}) \psi  ~~~{\rm
and} ~~~   \delta \bar \psi =i \alpha \bar \psi \gamma_5 ~.~ 
\end{equation}
 The generators of the transformation in Eq.(\ref{eqn:chrLR}) satisfy $\gamma_5^2 =1$ 
and $ \hat\gamma_5^2 \equiv  [\gamma_5 ( 1 - a D_{ov})]^2 =1 $.   One
therefore defines the left-right projections for quark fields as $\psi_L =
(1-\hat \gamma_5) \psi/2$ and $\psi_R = (1+ \hat \gamma_5) \psi/2$, leaving the
antiquark field decomposition as in the continuum.  Such a decomposition
is commonly done for writing chiral gauge theories~\cite{chgu} on the lattice
and is possible since $\psi$ and $\bar \psi$ are independent fields in the
Euclidean field theory.   In analogy with the Eq.(\ref{eqn:qcdlr}) of continuum
QCD, the action for the overlap quarks in presence of nonzero chemical
potential may now be written down as
\begin{eqnarray} 
S &=& \sum_{n,m}  [\bar \psi_{n,L} (aD_{ov} + a\mu
\gamma^4)_{nm} \psi_{m,L}+\bar \psi_{n,R} (aD_{ov} + a\mu
\gamma^4)_{nm} \psi_{m,R}]  \\  &=& \sum_{n,m}  \bar \psi_n [( 1- a\mu 
\gamma^4/2)aD_{ov} + a\mu \gamma^4]_{nm} \psi_m  ~. 
\label{eqn:ovelr} 
\end{eqnarray}  
It is easy to verify that this action is invariant under the chiral
transformation Eq.(\ref{eqn:chrLR}) for all values of $a\mu$ and $a$ and it
reproduces the continuum action in the limit of $a \to 0$ with $\mu \to \mu/M$
scaling.  In order to obtain the order parameter for checking if the symmetry
is spontaneously broken, one usually adds a linear symmetry breaking term.
Adding a quark mass term, $am (\bar \psi_R \psi_L + \bar \psi_L \psi_R)$, one
obtains the order parameter valid for all $T$ and $\mu$ on the lattice by
taking a derivative of the log of the partition function with respect to $am$
as, 
\begin{equation} 
\label{eqn:ppbar}
\langle \bar \psi \psi \rangle = \lim_{am \to 0} \lim_{V \to \infty} 
\langle {\rm Tr~} {\frac {1} { a K_{ov} + am +
a\mu \gamma^4}  } \rangle~. 
\end{equation} 
where $K_{ov} =  D_{ov}(1 -a D_{ov}/2)^{-1} $, such that 
$\{\gamma_5, K_{ov} \} = 0$.
Although the discussion above is for a single flavour of quark, i.e, $U(1)_L
\times U(1)_R$ symmetry, its generalization to $N_f$ flavours is
straightforward.  Indeed, since it relies only on the spin-structure, the
flavour index as well as the corresponding generator matrices just carry
through.

\section{Physical Picture}

Domain wall fermions are known to be akin to the the overlap fermions,
and may provide a physical picture for introduction of $\mu$ in the
way above.   Moreover, full QCD simulations with them are easier, so it is
also practically useful to check how the above exact chiral invariance 
can be preserved.  The domain wall action of ~\cite{shamir} for $\mu=0$, is 
\begin{eqnarray}
\label{eqn:dwacmuzero} 
S &=& \sum_{x,x'}\sum_{s,s'=1}^{N_5} a^4\bar \psi(x,s)\left[am~ 
\delta_{x,x'}\left(\delta_{s,1}\delta_{s',N_5}P_+ +
P_-\delta_{s,N_5}\delta_{s',1}\right)\right. \\ \nonumber
&& \left.- \delta_{x,x'}
\left(P_-\delta_{s',s+1}+P_+\delta_{s',s-1}\right)
+ \left(a_5 D_W(x,x')+ \delta_{x,x'}\right)\delta_{s,s'}   
\right]\psi(x',s')~, 
\end{eqnarray}
where $P_\pm = (1 \pm \gamma_5)/2$ and $N_5$, $a_5$ are
the number of sites and the lattice spacing in the fifth direction
respectively.  The physically relevant 4D  fermion field is
identified  with the fermion fields at the boundaries of the fifth dimension
as, 
\begin{equation} 
\Psi(x)=P_- \psi(x,1)+P_+ \psi(x,N_5)~~,~~\bar\Psi(x)=\bar \psi(x,1) P_+ 
+\bar \psi(x,N_5)P_-~. 
\end{equation}
These are the lowest energy field configurations of the five dimensional theory
and these correspond to the four dimensional chiral fermions in the limit
$m\rightarrow 0$ and $N_5\rightarrow\infty$.  In order to get a four
dimensional fermion determinant representing the dynamics of the physical quark
fields, one has to integrate out all the fermion degrees of freedom from the
five dimensional action and remove the contribution of the higher energy
configurations. The latter are the bulk modes since these  are delocalized in
the fifth dimension even when $N_5\rightarrow\infty$. Integrating out the
fermion degrees of freedom is conveniently done by recasting the domain wall
action in terms of the  fields $\eta_i$ localized on four dimensional branes
existing at each  site $i$ along the fifth dimension~\cite{eh}.  
These are related to the 5D fermion fields in an asymmetric way.  The 
neighbouring pairs of $\eta$ are related by a transfer matrix, 
$T =(1+a_5H_W P_+)^{-1}(1-a_5 H_WP_-)$ with $H_W=\gamma_5 D_W$.
Integrating over the $\eta$ fields successively one obtains a determinant which
is a functional of this transfer matrix:   
\begin{equation} 
D^{(5)}(ma) = \det \left[P_- -ma P_+ -T^{-N_5} \left(P_+ -ma
P_- \right)\right].
\end{equation}
It has the contribution of all the fermions in the five dimensional theory.
Choosing $ma=1$, one obtains contribution from all but the lowest energy
configurations which is divided out to obtain physical quark determinant.
Taking first $a_5\rightarrow 0$ limit and then $N_5\rightarrow \infty$, the
ratio of determinants turns out to be the determinant of the traditional
overlap Dirac matrix.  

To get correct physics at finite $\mu$ one needs to construct 
the number density corresponding to the physical degrees of freedom. 
We proposed \cite{gs12} that the following term be added to the domain wall 
action in Eq. (\ref{eqn:dwacmuzero}) for nonzero density:
$$\sum_{x,x'}\sum_{s,s'=1}^{N_5} a^4\bar \psi(x,s) a \mu~\gamma_4~
\delta_{x,x'}\left(\delta_{s,1}\delta_{s',1}P_-+
P_+\delta_{s,N_5}\delta_{s',N_5}\right)\psi(x',s').$$
We then follow the same procedure outlined above for the $\mu=0$ case and
rewrite the full action in terms of the $\eta$-fields.
The $\mu$-dependent term then becomes as below:
\begin{equation}
 \mu a\left[\bar\eta_1(a_5 H_W P_- -1)^{-1}\gamma_4P_-\eta_1
-\bar\eta_{N_5}(a_5 H_W P_- -1)^{-1}\gamma_4P_+\eta_1\right]
\end{equation}
It is important to note that unlike the Bloch-Wettig case~\cite{bwdw}, the
chemical potential in our proposal is coupled only to the physical fermions
which are strongly localized on the boundaries of the fifth dimension, hence
our transfer matrix continues to remain the same, i.e., is $\mu$-independent.
Moreover since we did not couple $\mu$ to the number density of the bulk modes,
the pseudo-fermion action remains the same as in the $\mu=0$ case.  Integrating
over the $\eta$ fields one obtains the physical fermion determinant in a
ratio form similar to that for $\mu=0$ with the $\det D^{(5)}(ma)$ in the
numerator generalised to, 
\begin{eqnarray} 
\nonumber
\det D^{(5)}(ma,\mu a)
&=&\det \left[P_- -ma P_+
+ a\mu \left(a_5 H_W P_- -1\right)^{-1} \gamma_4 P_-\right.\\
& &~~~~~-\left.T^{-N_5} \left(P_+ -ma
P_- - a\mu\left(a_5 H_W P_+ +1\right)^{-1}\gamma_4 P_+\right)\right].
\end{eqnarray}
and the pseudo-fermion determinant in the denominator remaining unchanged. On
taking the limits $a_5\rightarrow0$ and $N_5\rightarrow\infty$, the ratio of
determinant simplifies to $ \det [D_{ov} + (1- D_{ov}/2) (ma +  a \mu \gamma^4
) ]$, where both the dimensional parameters $\mu$ and $m$ have been scaled by a
factor of $1/M$ due to the tree level wavefunction renormalization~\cite{ehn}.
A little algebra shows that $\gamma^4$ can be commuted through in the
determinant above to yield the same overlap matrix of Eq. (\ref{eqn:ovelr})
with exact chiral symmetry on the lattice.

The action in Eq.(\ref{eqn:ovelr}) leads to an overlap fermion determinant
which is identical to that in the recent work~\cite{ns} with fermionic sources
in the overlap formalism of ~\cite{NeuNar}.  The main difference is, however,
in the necessity of sources in ~\cite{ns} to define the chiral symmetry.
Indeed, the chiral symmetry transformation there is local, defined as rotation
of the sources, while our Eq.(\ref{eqn:chrLR}) is nonlocal, defined as the
rotation of quark fields.  The left-right symmetry is in-built in the formalism
there whereas we needed to introduce the left-right projections in form of $L$-
and $R$-fields to do so.    

It is easy to verify that our $D_{ov}(a\mu)$, defined above, is not
$\gamma_5$-hermitian.  In general, therefore, it is not clear whether it may be
diagonalizable.  Noting that an $M$-scaling was essential in the continuum
limit of our $D_{ov}$ with $a \mu/M$ governing the density dependent term, it
is easily seen that for any arbitrary value of $\mu$ the density term can be
made small by choosing small enough $a$. Hence it suffices to study the chiral
anomaly at ${\cal O}((a \mu/M)^2)$. The leading term in such an expansion in $a
\mu/M$, is $D_{ov}(0)$ which is $\gamma_5$-hermitian.  Its eigenvalues come in
complex pairs $(\lambda, \lambda^*)$, with the corresponding eigenvectors
related by a $\gamma_5$-rotation.  Using these properties, we showed that
chiral anomaly arises as usual, and is governed by the number of zero modes of
$D_{ov}(0)$ even on the lattice.  In particular, the contribution of the
eigenvectors for the eigenvalue two, i.e, the `doublers', to chiral anomaly is
identically zero while that of the pairs $(\lambda, \lambda^*)$ cancels
mutually.  Using first order perturbation theory in $a \mu/M$, it is easy to
show that that the latter statement remain unchanged even for nonzero $\mu$ on
a fine enough lattice for our proposal while the former is true on any $a
\mu/M$.  Since zero modes continue to be eigenvalues of $\gamma_5$, all one
needs to do is to compute the trace of $\gamma_5$ in a new basis, rotated by a
unitary transformation.  After some algebra, one can show that the chiral
anomaly remains unchanged on lattice for arbitrary $\mu$ for our proposal to
introduce chemical potential for the overlap quarks, exactly as in the
continuum.

\section{Summary} 

In conclusion, we used the analogy with continuum QCD to demand a manifest
chiral invariance in terms of $L$ and $R$-fields on the lattice to obtain
$aD_{ov} + a\mu \gamma^4 (1- aD_{ov}/2)$ as the exact chiral invariant form of
the overlap Dirac matrix for any $\mu$ and/or lattice spacing.  We wrote down
the exact chiral order parameter on the lattice for chiral symmetry restoration
in the entire $T-\mu_B$ plane.  Chiral anomaly remains unaffected for our
proposal for small enough lattice spacing $a$, just as in continuum.  We
provided a physically appealing interpretation of our proposal by explicitly
writing down the corresponding domain wall fermion action at nonzero chemical
potential and taking the limit of infinite fifth dimension.


\end{document}